\documentclass[pra,twocolumn,showpacs]{revtex4}
\usepackage{amsmath}
\usepackage{graphicx}
\usepackage[usenames]{color}
\usepackage{amssymb}
\usepackage[dvips]{epsfig}

\newcommand{\beq}{\begin{equation}}
\newcommand{\eeq}{\end{equation}}
\newcommand{\beqa}{\begin{eqnarray}}
\newcommand{\eeqa}{\end{eqnarray}}
\newcommand{\ba}{\begin{array}}
\newcommand{\ea}{\end{array}}

\begin{document}

\title{Self-trapping of Fermi and Bose gases under spatially modulated
repulsive nonlinearity and transverse confinement}
\author{Luis E. Young-S.$^{1}$, L. Salasnich$^{2}$, and Boris A. Malomed$%
^{3} $}
\affiliation{$^{1}$Instituto de F\'{\i}sica Te\'orica, UNESP - Universidade Estadual
Paulista, 01.140-070 S\~ao Paulo, S\~ao Paulo, Brazil \\
$^2$Dipartimento di Fisica e Astromia ``Galileo Galilei'' and CNISM,
Universit\`a di Padova, Via Marzolo 8, 35131 Padova, Italy \\
$^{3}$Department of Physical Electronics, School of Electrical Engineering,
Faculty of Engineering, Tel Aviv University, Tel Aviv 69978, Israel}

\begin{abstract}
We show that self-localized ground states can be created in the
spin-balanced gas of fermions with repulsion between the spin components,
whose strength grows from the center to periphery, in combination with the
harmonic-oscillator (HO) trapping potential acting in one or two transverse
directions. We also consider the ground state in the non-interacting Fermi
gas under the action of the spatially growing tightness of the one- or
two-dimensional (1D or 2D) HO confinement. These settings are considered in
the framework of the Thomas-Fermi-von Weizs\"{a}cker (TF-vW) density
functional. It is found that the vW correction to the simple TF
approximation (the gradient term) is nearly negligible in all situations.
The properties of the ground state under the action of the 2D and 1D HO
confinement with the tightness growing in the transverse directions is
investigated too for the Bose-Einstein condensate (BEC) with the
self-repulsive nonlinearity.
\end{abstract}

\maketitle

\section{Introduction}

One of fundamental conclusions produced by ongoing studies of the dynamics
of matter waves in Bose-Einstein condensates (BECs) is that the effective
local nonlinearity, induced by inter-atomic collisions, gives rise to robust
bright solitons~and soliton complexes \cite%
{chap01:njp2003b,kono,fatk,Torner,Morsch,extra-reviews,boris}. In addition
to their fundamental significance, the solitons may be employed in
applications. In particular, the use of solitons in matter-wave
interferometers should help to dramatically increase the accuracy of these
devices, see Refs. \cite{Rosanov}-\cite{Lev} and a recent review~\cite%
{billam2}.

In the usual settings, the existence of bright solitons requires the
presence of the self-focusing nonlinearity. They may also be supported by
nonlinear lattices, which feature alternation of spatial domains of
self-focusing and defocusing, which give rise to the corresponding periodic
\textit{pseudo-potential} \cite{boris,sala-nl}. On the other hand,
self-defocusing nonlinearities, acting in the combination with periodic
linear (lattice) potentials, support bright solitons of the bandgap type
\cite{kono,fatk,Morsch}. Nevertheless, a common belief was that the
self-defocusing per se could not give rise to bright solitons. The situation
had changed when it was demonstrated that the repulsive cubic nonlinearity
with the local strength growing with the distance from the center, $r$,
faster than $r^{D}$, where $D$ is the spatial dimension, can readily support
stable solitons and solitary vortices \cite{Barcelona,Barcelona2}. In the
same works, it was proposed how the corresponding profiles of the
nonlinearity modulation can be created in BEC and nonlinear optics. In
particular, the spatial modulation of the scattering length of inter-atomic
collisions, induced by spatially inhomogeneous magnetic or optical fields
via the Feshbach-resonance mechanism, can give rise to the required profile
in BEC. Similarly, robust solitons and vortices were predicted in a more
exotic model, with the constant strength of the defocusing nonlinearity but
the diffraction coefficient decaying faster than $r^{-D}$ \cite{Zhong}.

It may be interesting to implement this new option for the creation of
bright solitons by purely repulsive nonlinearity landscapes in other
physical settings. For instance, it was recently demonstrated that the same
mechanism works too in one- and two-dimensional (1D and 2D) media with the
repulsive quintic nonlinearity \cite{Zeng}.

The present work aims to propose a way of creating bright solitons in Fermi
gases, with balanced spin-up and spin-down components. The possibility is to
combine the spatial growth of the \textit{s}-wave scattering length, which
accounts for the repulsion between the components, in one or two directions,
and the harmonic-oscillator (HO)\ confining potential acting in the other
directions. It should be noted that the dilute two-spin-component fermionic
gases with the repulsive \textit{s}-wave interaction do not feature phase
coherence, as they are not superfluids \cite{lipparini}. Nevertheless, their
ground states can be accurately described by means of the density-functional
theory \cite{hkt,tosi,Brandon}.

Here we adopt the Thomas-Fermi-von Weizs\"{a}cker (TF-vW) form of the
density functional to predict density profiles and chemical potentials of
the Fermi droplets (i.e., bright solitons) in two different configurations,
which are considered in Sections II and III, respectively: the quadratic
growth of the scattering length in one direction ($z$) and transverse HO
confinement in the $\left( x,y\right) $ plane; or the quadratic growth of
the scattering length in the $\left( x,y\right) $ plane and HO confinement
acting along $z$. Then, in Sections IV\ and V, following previous results
obtained in the framework of the mean-field description of BEC \cite%
{sala-mod,we,Canary}, we consider Fermi gases without intrinsic interactions
(in particular, it may be a spin-polarized, i.e., single-spin-component,
gas) and demonstrate that they give rise to localized ground states under
the action of the 2D or 1D HO confinement, if its tightness, i.e., the
corresponding confinement frequency, grows in the transverse directions
faster than $\sqrt{|z|}$ or $r^{4}$, respectively. In addition, we
demonstrate that the same confinements with the spatially growing tightness
support, in a similar way, localized ground states in BEC with the
self-repulsive nonlinearity.

\section{1D nonlinearity modulation with the 2D transverse confinement for
the spin-balanced interacting Fermi gas}

In the presence of an external trapping potential $U_{\mathrm{ext}}(x,y,z)$,
the Hohenberg-Kohn theorem \cite{hkt} ensures that the single-body local
density $n(x,y,z)$ of the ground state of a quantum system composed of
interacting identical particles can be obtained by minimizing the energy
functional,
\begin{equation}
E[n]=F[n]+\int \int \int U_{\mathrm{ext}}(x,y,z)\ n(x,y,z)\ dxdydz,
\label{eq:ETF}
\end{equation}%
where $F[n]$ is an internal-energy functional, which is independent of $U_{%
\mathrm{ext}}(x,y,z)$. For the dilute normal Fermi gas with two
equally-populated spin states, the TF-vW internal energy at zero-temperature
is written as \cite{lipparini}%
\begin{eqnarray}
F[n] &=&\int \int \int {\ }\left[ {\frac{3}{5}}{\frac{\hbar ^{2}}{2m}}(3\pi
^{2})^{2/3}n^{5/3}+\lambda {\frac{\hbar ^{2}}{8m}}{\frac{(\nabla n)^{2}}{n}}%
\right.   \notag \\
&&\left. +{\frac{1}{4}}g(z)n^{2}\right] dx\,dy\,dz,  \label{E}
\end{eqnarray}%
with the nonlinearity strength determined by the \textit{s}-wave
fermion-up--fermion-down scattering length, which (as we assume in this
work) may be modulated along the axial direction, $a_{\uparrow \downarrow
}(z)$:
\begin{equation}
g(z)=\left( 4\pi \hbar ^{2}{/m}\right) a_{\uparrow \downarrow }(z).
\label{g}
\end{equation}%
The first term in functional (\ref{E}) is the TF kinetic energy of the Fermi
gas at zero temperature, while the second term is the vW correction to the
kinetic energy of the inhomogeneous gas (the surface term). For the
coefficient in front of the vW term, the phenomenological value, $\lambda
=1/3$, may be adopted for non-superfluid fermions \cite{tosi}. The third
term in (\ref{E}) is the mean-field approximation of the \textit{s}-wave
interaction between fermions with opposite spins.

The HO confinement in the transverse plane is accounted for by the external
potential,
\begin{equation}
U_{\mathrm{ext}}(x,y)={(1/2)}m\omega _{\bot }^{2}r^{2},  \label{U2D}
\end{equation}%
with $r=(x^{2}+y^{2})^{1/2}$ and confinement radius $a_{\bot }=\sqrt{\hbar
/(m\omega _{\bot })}$. We suppose that the spatially modulated scattering
length also features the quadratic coordinate dependence, with a
characteristic scale $a_{0}>0$,
\begin{equation}
a_{\uparrow \downarrow }(z)=\left( a_{0}/a_{\bot }^{2}\right) \ z^{2}.
\label{a1D}
\end{equation}

By minimizing the energy functional (\ref{E}), which is subject to the
normalization constraint, imposed by the fixed total number of fermions,%
\begin{equation}
N=\int \int \int n(x,y,z)\ dxdydz,  \label{norma}
\end{equation}%
one derives the equation for the local density,
\begin{equation}
\left[ -{\lambda }\nabla ^{2}+(3\pi ^{2})^{2/3}n^{2/3}+r^{2}+{\gamma }z^{2}n%
\right] \sqrt{n}=2\mu \sqrt{n},  \label{eq}
\end{equation}%
with chemical potential $\mu $ fixed by normalization (\ref{norma}).
Equation (\ref{eq}) is written in the scaled notation, with lengths measured
in units of $a_{\bot }$ and energies in units of $\hbar \omega _{\bot }$. In
this form, the equations depend on the single parameter, \textit{viz}., the
adimensional interaction strength,
\begin{equation}
\gamma \equiv 4\pi a_{0}/a_{\bot }.  \label{gamma}
\end{equation}%
Multiplying the left-hand side of Eq.(\ref{eq}) by $\sqrt{n}$ and
integrating the result over the spatial coordinates, we obtain an expression
for the chemical potential in terms of density $n(x,y,z)$:
\begin{eqnarray}
\mu  &=&{\frac{1}{2N}}\int \int \int \left[ -{\lambda }\sqrt{n}\nabla
^{2}\left( \sqrt{n}\right) +(3\pi ^{2})^{2/3}\ n^{5/3}\right.  \\
&&\left. +r^{2}n+{\gamma }z^{2}n^{2}\right] dx\,dy\,dz.
\end{eqnarray}

In the TF regime, when the vW correction is negligible, Eq. (\ref{eq}) for
the density profile reduces to a cubic algebraic equation for $n^{1/3}$,
\begin{equation}
(3\pi ^{2})^{2/3}n^{2/3}+r^{2}+{\gamma }z^{2}n=2\mu ,  \label{eq-tf}
\end{equation}%
at $r<r_{\mathrm{TF}}$, and $n=0$ at $r>r_{\mathrm{TF}}$, where the TF
radius is
\begin{equation}
r_{\mathrm{TF}}=\sqrt{2\mu }.  \label{rTF}
\end{equation}

\begin{figure}[tbp]
\begin{center}
{\includegraphics[width=9.cm,clip]{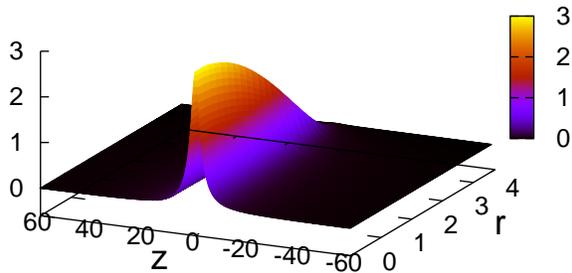}}
\end{center}
\caption{(Color online). The three-dimensional density profile $n(r,z)$ of
the two-component spin-balanced Fermi gas, under the action of the 1D
nonlinearity modulation, see Eq.\ (\protect\ref{a1D}), and the
harmonic-oscillator confinement in the transverse plane [Eq. (\protect\ref%
{U2D})]. The chemical potential is $\protect\mu =10$, adimensional
nonlinearity strength $\protect\gamma =1$ [see Eq. (\protect\ref{gamma})],
and $\protect\lambda =1/3$.}
\label{fig1}
\end{figure}

To find solutions of full equation (\ref{eq}), including the vW correction,
an imaginary-time-derivative term was added to it, and ensuing stationary
solutions were obtained by means of the finite-difference Crank-Nicolson
predictor-corrector method, which keeps the fixed value of $N$ \cite%
{sala-numerics}. In Fig. \ref{fig1} we plot the so obtained 3D density
profile $n(r,z)$ of the spin-balanced Fermi gas, produced by the numerical
solution of Eq. (\ref{eq}) for $\mu =10$, the nonlinearity strength $\gamma
=1$, and $\lambda =1/3$ (as said above), which corresponds to the
self-trapped mode built of $N=612$ fermions. The figure shows that, while
along the $r$ direction the density practically vanishes at $r_{\mathrm{TF}}=%
\sqrt{20}\simeq 4.47$, as predicted by the TF approximation, along the $z$
direction the density vanishes only at $|z|\rightarrow \infty $.

\begin{figure}[tbp]
\begin{center}
{\includegraphics[width=8.cm,clip]{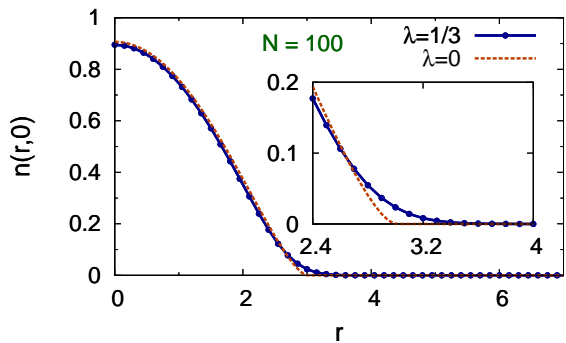}}
\end{center}
\caption{(Color online). Radial density $n(r,0)$ of the spin-balanced Fermi
gas with nonlinearity strength $\protect\gamma =1$ and number of particles $%
N=100$, in the same setting as in \protect\ref{fig1}. Solid lines: solutions
of the full equation (\protect\ref{eq}), which includes the von Weizs\"{a}%
cker term with $\protect\lambda =1/3$. Dashed lines: solutions produced by
the Thomas-Fermi (TF) approximation based on Eq. (\protect\ref{eq-tf})
[i.e., Eq. (\protect\ref{eq}) with $\protect\lambda =0$]. Insets display a
blowup of the structure of each state across the surface layer.}
\label{fig2}
\end{figure}

In Fig. \ref{fig2} we plot the radial density, $n(r,z=0)$, for $N=100$
fermions, and compare the solutions of the full equation (\ref{eq}) with the
TF approximation based on Eq.(\ref{eq-tf}). The figure shows that the vW
gradient term mainly affects the behavior of the density near the surface
layer (see the insets in the figure): instead of vanishing at a finite
distance $r_{\mathrm{TF}}$ from the center, the density vanishes at $%
r\rightarrow \infty $. Nevertheless, this effect is weak, and it becomes
negligible for larger $N$.

\begin{figure}[tbp]
\begin{center}
{\includegraphics[width=8.cm,clip]{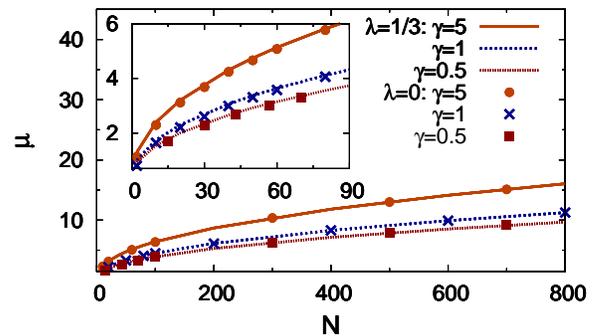}}
\end{center}
\caption{(Color online). Chemical potential $\protect\mu $ (in units of $%
\hbar \protect\omega _{\perp }$) vs. the number of particles, $N$, for the
spin-balanced Fermi gas with nonlinearity strengths $\protect\gamma =5,\ 1$
and $0.5$, in the same settings as in Figs. \protect\ref{fig1} and \protect
\ref{fig2}. Lines represent the dependences obtained from the full
Thomas-Fermi-von Weizs\"{a}cker model, i.e., Eq. (\protect\ref{eq}) with $%
\protect\lambda =1/3$, while points correspond to the TF approximation ($%
\protect\lambda =0$). The inset is a blowup of the dependence at small
values of $N$. }
\label{fig3}
\end{figure}

Similar features are exhibited by the dependence of the chemical potential, $%
\mu $, on the number of atoms, $N$, as shown in Fig. \ref{fig3}.\ The change
of these dependences caused by the vW term is very small even for the state
built of a dozen of atoms, see the inset in the figure. Thus, the simplified
TF functional, disregarding the vW corrections ($\lambda =0$), is sufficient
for the study of the ground state in the present model.

\begin{figure}[tbp]
\begin{center}
{\includegraphics[width=8.cm,clip]{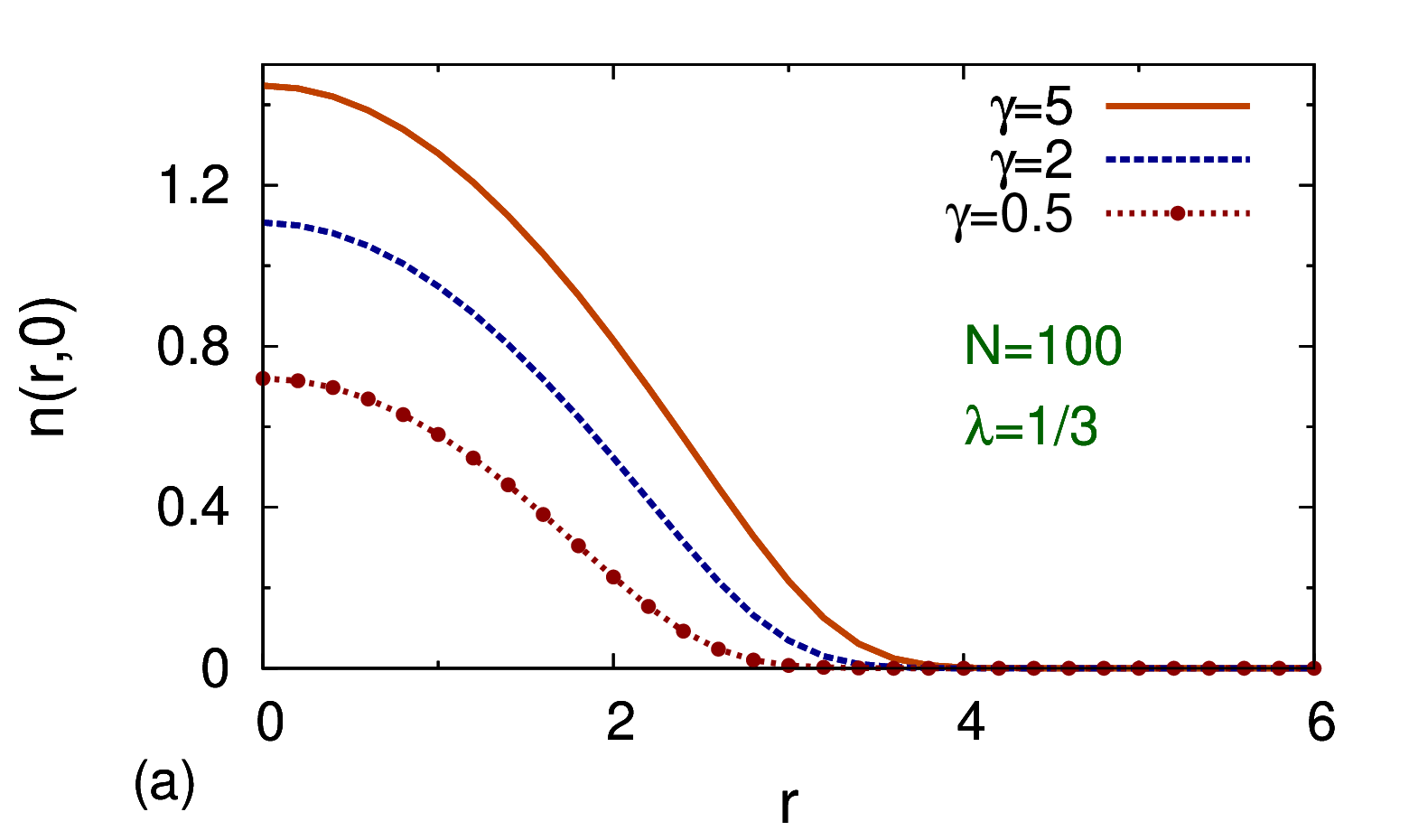}} {%
\includegraphics[width=8.cm,clip]{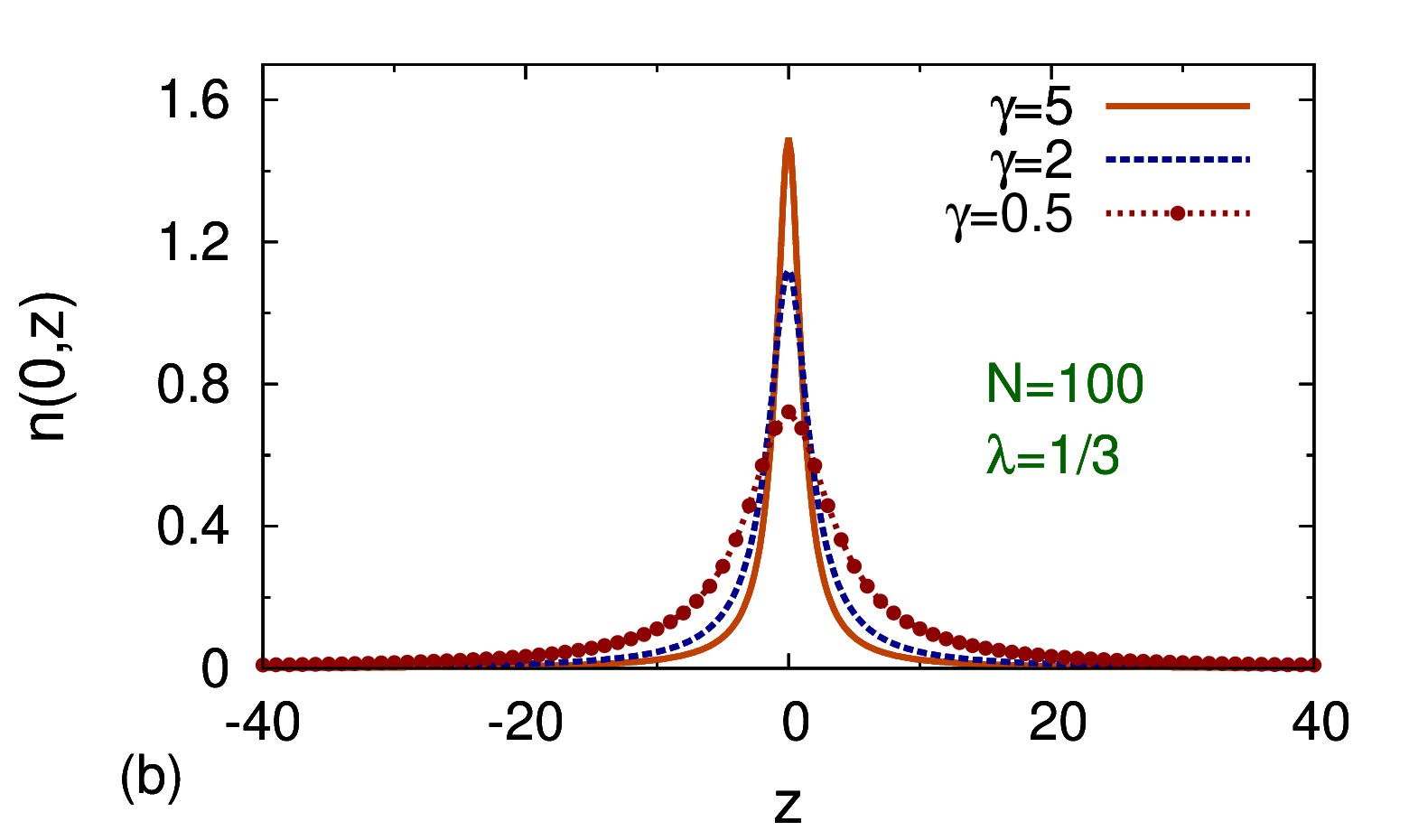}}
\end{center}
\caption{(Color online). (a) Radial $n(r,z=0)$ and (b) axial $n(r=0,z)$
density profiles of the ground state composed of $N=100$ fermions, for $%
\protect\lambda =1/3$ in Eq. (\protect\ref{eq}) and three values of
nonlinearity strength $\protect\gamma $, which is defined as per Eq. (%
\protect\ref{gamma}). The setting is the same as in Figs. \protect\ref{fig1}-%
\protect\ref{fig3}.}
\label{fig4}
\end{figure}

Regarding the ground state, it is natural to expect that its width reduces
with the growth of strength $\gamma $ of the self-repulsive nonlinearity.
This expectation is confirmed by Fig. \ref{fig4}. In the upper and lower
panels of this figure, we plot the radial and axial density profiles,
\textit{viz}., $n(r,z=0)$ and $n(r=0,z)$ respectively, for $N=100$, $\lambda
=1/3$ (i.e., the vW is taken into account here, although it produces very
little change) and three values of the nonlinearity strength, $\gamma =0.5$,
$2$, and $5$.

\begin{figure}[tbp]
\begin{center}
{\includegraphics[width=8.cm,clip]{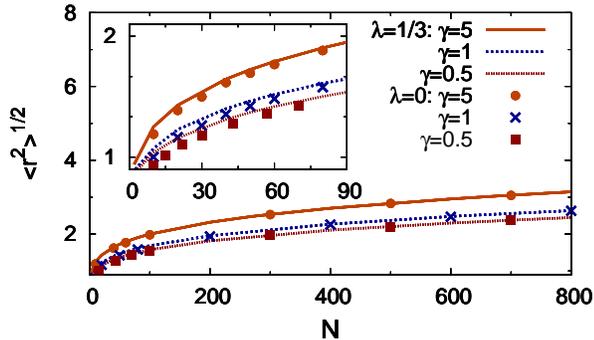}}
\end{center}
\caption{(Color online). The rms radial size $\langle r^{2}\rangle ^{1/2}$
vs. the number of fermions, $N$, for nonlinearity strengths $\protect\gamma %
=5,\ 1$ and $0.5$. Lines are produced by full equation (\protect\ref{eq})
with $\protect\lambda =1/3$, while points correspond to the TF
approximation, $\protect\lambda =0$. The inset shows the magnification of
the dependence at small values of $N$. The setting is the same as in Figs.
\protect\ref{fig1}-\protect\ref{fig4}. }
\label{fig5}
\end{figure}

As said above, chemical potential $\mu $ determines the total number $N$ of
atoms in the ground state, hence its width increases with $\mu $, for fixed
nonlinearity strength $\gamma $. In Fig. \ref{fig5} we plot the
root-mean-square (rms) radial size of the ground state, $\langle
r^{2}\rangle ^{1/2}$, versus $N$, for three values of the nonlinearity
strengths, $\gamma =5,\ 1$ and $0.5$, using the the full equation (\ref{eq})
with $\lambda =1/3$ (lines in Fig. \ref{fig5}) and the TF approximation
corresponding to $\lambda =0$ (points in Fig. \ref{fig5}). The results
produced by the full and simplified models are essentially the same for
different values of $\gamma $ and, virtually, for all $N$.

\section{2D nonlinearity modulation with the 1D transverse confinement}

We now consider the setting with the HO confinement acting along the
longitudinal direction, \textit{viz}.,
\begin{equation}
U_{\mathrm{ext}}(z)={(1/2)}m\omega _{z}^{2}z^{2},  \label{z}
\end{equation}%
cf. Eq. (\ref{U2D}). On the other hand, the spatial modulation of the
scattering length, $a_{\uparrow \downarrow }$, is adopted here to be
two-dimensional, cf. Eq. (\ref{a1D}):
\begin{equation}
a_{\uparrow \downarrow }(r)=\left( a_{0}/a_{z}^{2}\right) \ r^{2},
\label{up-down}
\end{equation}%
with $a_{0}>0$, where $a_{z}=\sqrt{\hbar /(m\omega _{z})}$ is the
characteristic length of the longitudinal HO confinement (\ref{z}). Thus,
the present setting is\ a reverse of that considered in the previous section.

The corresponding TF-vW internal energy is
\begin{eqnarray}
F[n] &=&\int \int \int {\ }\left[ {\frac{3}{5}}{\frac{\hbar ^{2}}{2m}}(3\pi
^{2})^{2/3}n^{5/3}+\lambda {\frac{\hbar ^{2}}{8m}}{\frac{(\nabla n)^{2}}{n}}%
\right.  \\
&&+\left. {\frac{1}{4}}g(r)n^{2}\right] \ dx\,dy\,dz,
\end{eqnarray}%
with $g(r)$ given by Eq. (\ref{g}), in which $a_{\uparrow \downarrow }(z)$
is replaced by $a_{\uparrow \downarrow }(r)$, see Eq. (\ref{up-down}). By
minimizing the full energy functional, $E[n]=F[n]+\int \int \int U_{\mathrm{%
ext}}(z)n(x,y,z)dxdydz$, with the constraint of the normalization of $n(%
\mathbf{r})$, we obtain
\begin{equation}
\left[ -{\lambda }\nabla ^{2}+(3\pi ^{2})^{2/3}n^{2/3}+z^{2}+{\Gamma }r^{2}n%
\right] \sqrt{n}=2\mu \sqrt{n},  \label{eqqq}
\end{equation}%
cf. Eq. (\ref{eq}). In Eq. (\ref{eqqq}) lengths are measured in units of $%
a_{z}$, and energies in units of $\hbar \omega _{z}$, which leaves, as the
single control parameter, the adimensional strength of the nonlinearity,%
\begin{equation}
\Gamma \equiv 4\pi a_{1}/a_{z},  \label{Gamma}
\end{equation}%
cf. Eq. (\ref{gamma}).

In the TF approximation, which, as before, neglects the vW gradient term,
the density profile, $n(r,z)$, satisfies an algebraic equation [cf. Eq. (\ref%
{eq-tf})]:
\begin{equation}
(3\pi ^{2})^{2/3}n^{2/3}+z^{2}+{\Gamma }r^{2}n=2\mu .  \label{eqq-tf}
\end{equation}%
Setting $n(r,z)=0$ in Eq.(\ref{eqq-tf}), one gets the TF axial size of the
ground state, $z_{\mathrm{TF}}=\sqrt{2\mu }$, with $n=0$ at $|z|>z_{\mathrm{%
TF}}$ in the TF approximation, cf. Eq. (\ref{rTF}).

\begin{figure}[tbp]
\begin{center}
{\includegraphics[width=8.5cm,clip]{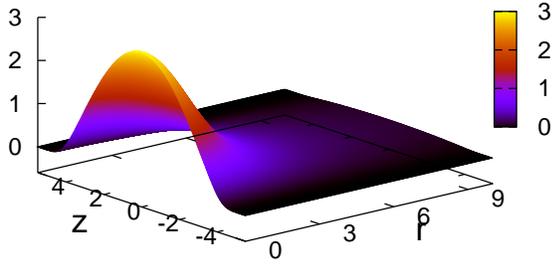}}
\end{center}
\caption{(Color online) Three-dimensional density profile $n(r,z)$ of the
two-component spin-balanced Fermi gas under the 1D external confinement, see
Eq. (\protect\ref{z}), and 2D modulation of the local nonlinearity in the
transverse plane [Eq. (\protect\ref{up-down})]. The chemical potential is $%
\protect\mu =10$, adimensional nonlinearity strength is $\Gamma =2$, and $%
\protect\lambda =1/3$ is adopted in Eq. (\protect\ref{eqqq}). Lengths are
measured in units of $a_{z}$, and energies in units of $\hbar \protect\omega %
_{z}$.}
\label{fig6}
\end{figure}

In Fig. \ref{fig6} the 3D density profile $n(r,z)$ of the Fermi gas is
plotted for $\mu =10$, $\Gamma =2$, and $\lambda =1/3$ in Eq. (\ref{eqqq}).
The figure shows that, while along the $z$ direction the density practically
vanishes at $z_{\mathrm{TF}}=\sqrt{20}\simeq 4.47$, along $r$ the density
vanishes only at $r\rightarrow \infty $.

\begin{figure}[tbp]
\begin{center}
{\includegraphics[width=8.cm,clip]{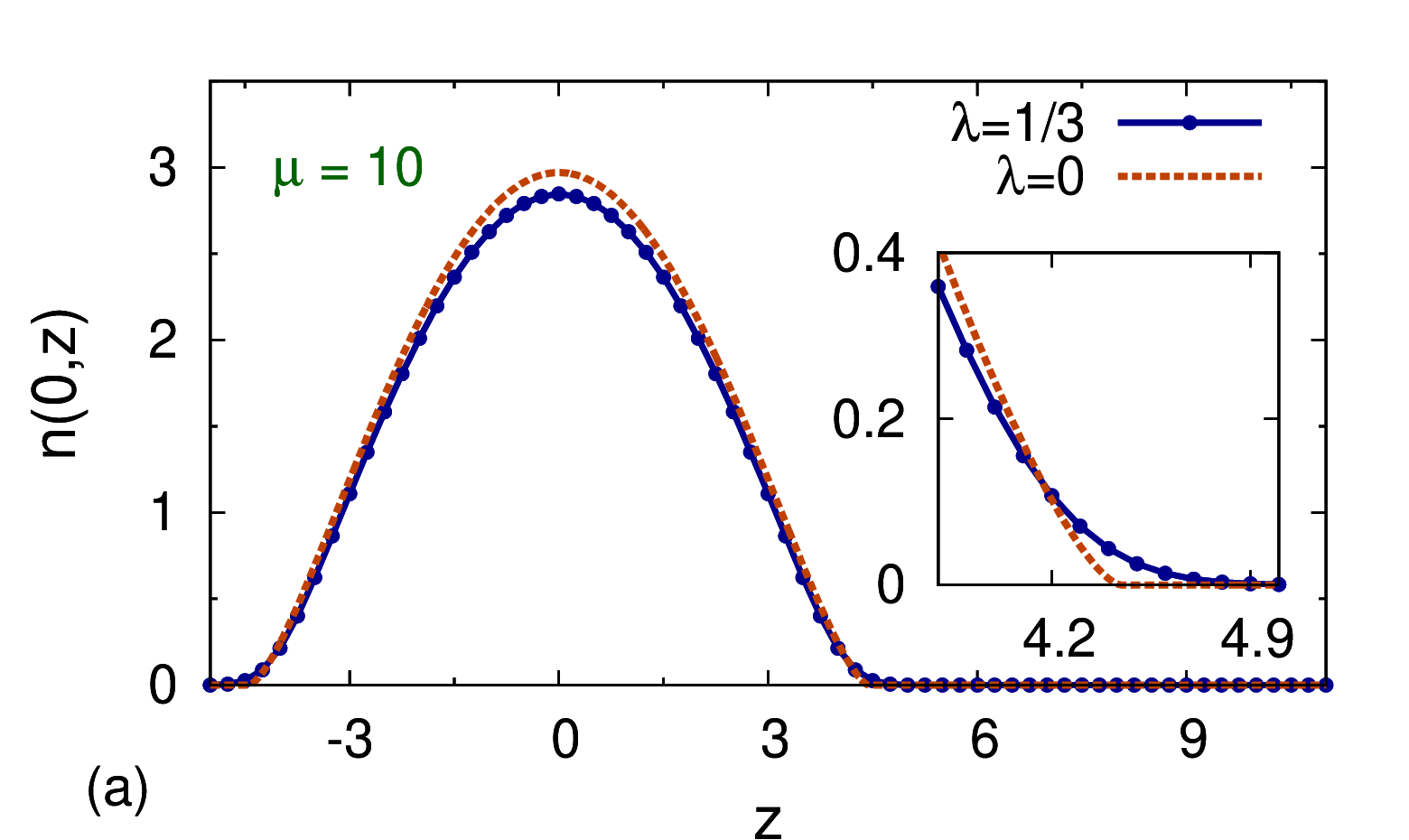}} {%
\includegraphics[width=8.cm,clip]{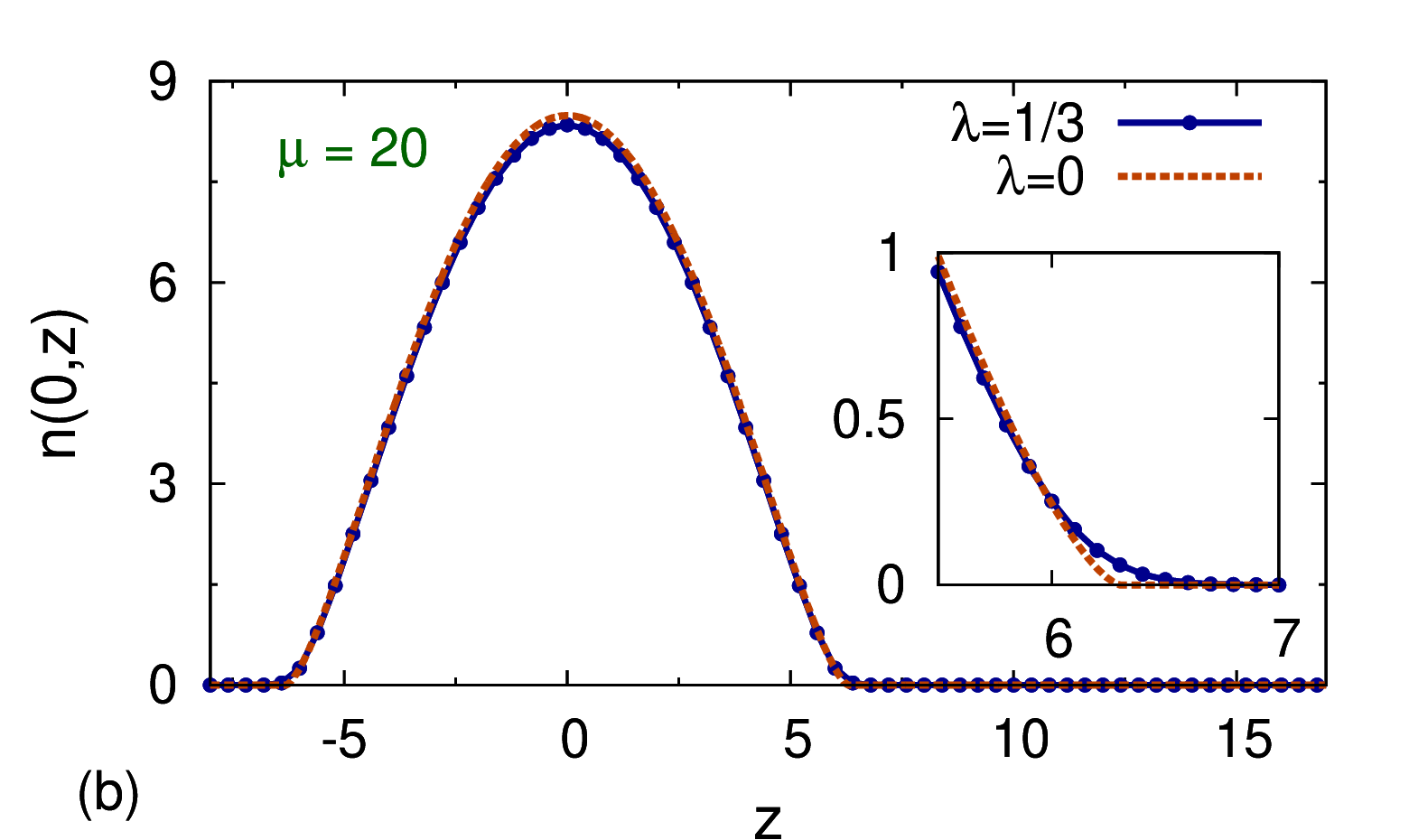}}
\end{center}
\caption{(Color online). Axial density profiles $n(0,z)$ of the
two-component spin-balanced Fermi gas under the 1D confinement [Eq. (\protect
\ref{z})] and 2D modulation of the nonlinearity [Eq. (\protect\ref{up-down}%
)], whose global strength is $\Gamma =2$,[see Eq. (\protect\ref{Gamma}), for
two values of the chemical potential: $\protect\mu =10$ (a) and $\protect\mu %
=20$ (b). Solid lines: solutions of Eq. (\protect\ref{eqqq}) with $\protect%
\lambda =1/3$; dashed lines: solutions of Eq. (\protect\ref{eqq-tf}), i.e.,
Eq. (\protect\ref{eqqq}) with $\protect\lambda =0$. Units are the same as in
Fig. \protect\ref{fig6}.}
\label{fig7}
\end{figure}

In Fig. \ref{fig7} we plot axial density $n(0,z)$ for two values of the
chemical potential, $\mu =$ $10$ and $20$, in panels (a) and (b),
respectively, both taken with $\Gamma =2$. In this figure we compare
solutions of the full equation (\ref{eqqq}) to their counterparts produced
by the TF approximation, which is based on Eq. (\ref{eqq-tf}). The figure
shows, once again, that the vW gradient term mainly affects the behavior of
the density in the surface layer: Instead of vanishing at a finite distance (%
$z_{\mathrm{TF}}$) from the center of the cloud, it vanishes at $%
|z|\rightarrow \infty $. However, this effect is weak, becoming negligible
at larger values of the chemical potential.

\begin{figure}[tbp]
\begin{center}
{\includegraphics[width=8.cm,clip]{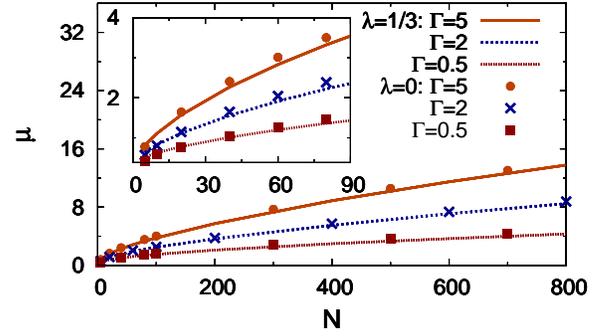}}
\end{center}
\caption{(Color online). Chemical potential $\protect\mu $ versus number $N$
of fermionic atoms for the two-component spin-balanced Fermi gas under the
1D confinement and 2D modulation of the local nonlinearity. The nonlinearity
strength takes values $\Gamma =5,\ 2$ and $0.5$. Lines are obtained from
full equation (\protect\ref{eqqq}), while points correspond to the TF
approximation ($\protect\lambda =0$). Energies are in units of $\hbar
\protect\omega _{z}$.}
\label{fig8}
\end{figure}

As shown in Fig. \ref{fig8}, a similar phenomenon is observed in the
dependences of the chemical potential $\mu $ on $N$, which were produced by
the full equation \ref{eqqq} with $\lambda =1/3$ (lines in Fig. \ref{fig8}),
and by the TF approximation (\ref{eqq-tf}) (points in Fig. \ref{fig8}). The
effect of the vW gradient term remains inconspicuous even at small values of
$N$, and at all values of $\Gamma $.

\begin{figure}[tbp]
\begin{center}
{\includegraphics[width=8.cm,clip]{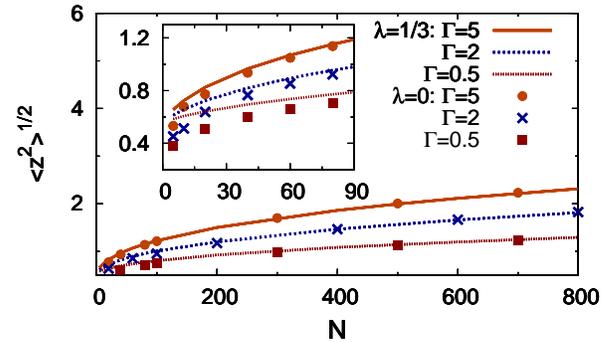}}
\end{center}
\caption{(Color online). The rms size $\langle z^{2}\rangle ^{1/2}$ vs.
number $N$ of fermions of the ground state with nonlinearity strengths $%
\Gamma =5,\ 2$ and $0.5$. Lines and points represent the results obtained,
respectively, from the full equation (\protect\ref{eqqq}) with $\protect%
\lambda =1/3$, and from the TF approximation corresponding to $\protect%
\lambda =0$. The inset shows details of the dependences at smaller values of
$N$. The setting is the same as in Figs. \protect\ref{fig6}-\protect\ref%
{fig8}.}
\label{fig9}
\end{figure}

A natural expectation that the radial width of the ground state reduces with
the increase of the nonlinearity strength, $\Gamma $ [i.e., the gas is
stronger compressed by the effective nonlinear (pseudo)potential], is
confirmed by Fig. \ref{fig9}. In this figure we plot the rms size of the
ground state, $\langle z^{2}\rangle ^{1/2}$, versus the number of particles,
$N$, for three values of the nonlinearity strength, $\Gamma =5,\ 2$ and $0.5$%
, using the full equation (\ref{eqqq}) with $\lambda =1/3$, the TF
approximation corresponding to $\lambda =0$ (lines and points, respectively,
in Fig. \ref{fig9}). Thus, we conclude that the effect of the vW term is
inessential in this setting too.

\section{Self-trapping in the non-interacting Fermi gas and self-repulsive
BEC due to the axial modulation of the 2D confinement}

\subsection{The non-interacting Fermi gas under the 2D confinement with the
strength growing in the axial direction}

In the spin-balanced \emph{ideal} Fermi gas, without any interaction between
the two spin components, or the single-component spin-polarized gas, in
which direct interactions are suppressed by the Pauli principle, it is
possible to induce self-trapping along the longitudinal direction ($z$) by
introducing a $z$-dependent modulation of the transverse HO confinement
frequency, $\omega _{\bot }$. The corresponding energy functional is
\begin{eqnarray}
E[n] &=&\int \int \int \left[ {\frac{3}{5}}{\frac{\hbar ^{2}}{2m}}(3\pi
^{2})^{2/3}n^{5/3}+\lambda {\frac{\hbar ^{2}}{8m}}{\frac{(\nabla n)^{2}}{n}}%
\right.  \\
&&\left. +{\frac{1}{2}}m\omega _{\bot }^{2}(z)(x^{2}+y^{2})\,n\right] \
dx\,dy\,dz\;.
\end{eqnarray}%
By minimizing this functional, subject, as before, to the constraint of the
normalization of $n(\mathbf{r})$, one obtains
\begin{equation}
\left[ -{\lambda }\nabla ^{2}+(3\pi ^{2})^{2/3}n^{2/3}+\omega _{\bot
}^{2}(z)r^{2}\right] \sqrt{n}=2\mu \sqrt{n},  \label{eqqm}
\end{equation}%
where $\mu $ is the chemical potential fixed by the total number of atoms, $N
$ [see Eq. (\ref{norma})]. Solving Eq.(\ref{eqqm}), we aim to produce the
longitudinal density profile,
\begin{equation}
n_{1}(z)\equiv \int \int n(\mathbf{r})\ dxdy.  \label{eq_n1}
\end{equation}

The TF approximation corresponds, as above, to setting $\lambda =0$ in Eq. (%
\ref{eqqm}), which makes it a simple algebraic equation that immediately
yields an explicit solution [this was not available in the presence of the
interaction between the spin components, cf. Eqs. (\ref{eq}) and (\ref{eq-tf}%
)]:
\begin{equation}
n\left( r,z\right) =\left\{
\begin{array}{c}
\left( 3\pi ^{2}\right) ^{-1}\left[ 2\mu -\omega _{\perp }^{2}(z)r^{2}\right]
^{3/2},~\mathrm{at}~~\omega _{\perp }^{2}(z)r^{2}<2\mu , \\
0,~\mathrm{at}~~\omega _{\perp }^{2}(z)r^{2}>2\mu
\end{array}%
\right.   \label{TF}
\end{equation}%
(obviously, the solution exists only for $\mu >0$). This solution is
physically meaningful if its norm converges, which implies the convergence
of $\int_{0}^{z}r_{\mathrm{TF}}^{2}\left( z^{\prime }\right) dz^{\prime
}\equiv 2\mu \int_{0}^{z}\omega _{\perp }^{-2}\left( z^{\prime }\right)
dz^{\prime }$ at $z\rightarrow \infty $. The eventual condition is that $%
\omega _{\perp }^{2}(z)$ must grow faster than $|z|$, i.e.,%
\begin{equation}
\omega _{\perp }(z)/\sqrt{|z|}\rightarrow \infty ,  \label{inf}
\end{equation}%
at $|z|\rightarrow \infty $. Further, the substitution of explicit solution (%
\ref{TF}) into Eq. (\ref{eq_n1}) makes it possible to obtain the TF
approximation for $n_{1}$ in an explicit form too:
\begin{equation}
n_{1}(z)={\frac{16\sqrt{2}}{15\pi }}\left( {\frac{m\mu }{\hbar ^{2}}}\right)
^{5/2}a_{\bot }^{4}(z)\;.  \label{eqqm2}
\end{equation}

\begin{figure}[tbp]
\begin{center}
{\includegraphics[width=1.02\linewidth,clip]{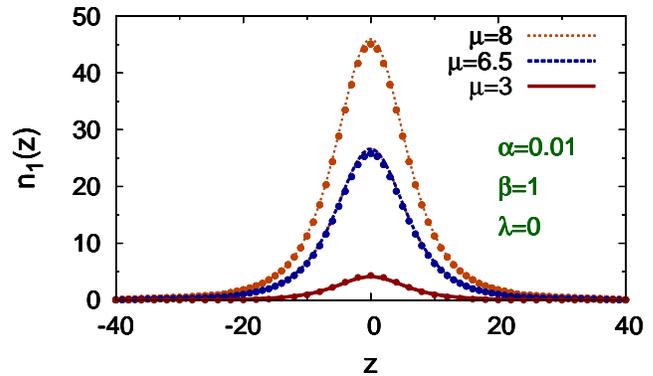}}
\end{center}
\caption{(Color online). Profiles of the longitudinal density $n_{1}(z)$ of
the non-interacting Fermi gas, with the $z$-dependent modulation of the
transverse-confinement frequency, chosen as per Eq. (\protect\ref{eq_wz})
with $\protect\alpha =0.01$ and $\protect\beta =1$, for three values of the
chemical potential: $\protect\mu =8,~6.5$, and $5$. Lines: a numerical
solution of Eq. (\protect\ref{eqqm}) for $\protect\lambda =0$. Points:
analytical result (\protect\ref{eqqm2}).}
\label{fig10}
\end{figure}

Below we consider the most natural modulation form,
\begin{equation}
\omega _{\bot }(z)=\alpha z^{2}+\beta ,  \label{eq_wz}
\end{equation}%
with positive $\alpha $ and $\beta $. Obviously, it satisfies condition (\ref%
{inf}). In Fig. \ref{fig10} we plot longitudinal profiles of the reduced
density (\ref{eq_n1}) for the $z$-dependent modulation of the
transverse-confinement strength given by Eq. (\ref{eq_wz}) with $\alpha
=0.01 $ and $\beta =1$. Dots in Fig. \ref{fig10} represent analytical result
(\ref{eqqm2}), and lines depict (for the sake of checking the correctness of
the analytical result) the numerical solution of Eq. (\ref{eqqm}) with $%
\lambda =0$, obtained as in Ref. \cite{sala-numerics} [in the latter case,
the 3D density is numerically integrated in the transverse plane to reduce
it to $n_{1}$, see Eq. (\ref{eq_n1}).

\begin{figure}[tbp]
\begin{center}
{\includegraphics[width=1.0\linewidth,clip]{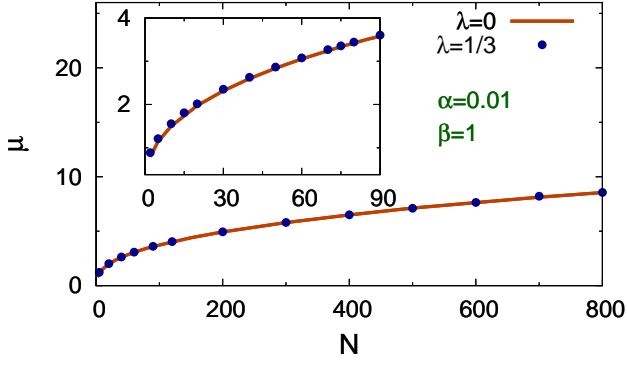}}
\end{center}
\caption{(Color online). Chemical potential $\protect\mu $ vs. the number of
fermionic atoms, $N$, of the ground state of the non-interacting Fermi gas
subject to modulation (\protect\ref{eq_wz}) of the frequency of the
transverse confinement. As indicated in the figures, the TF approximation ($%
\protect\lambda =0$) produces the dependence which is very close to that
generated by the full model, including the von Weizs\"{a}cker term ($\protect%
\lambda =1/3$). The inset is a blowup of the dependence at small values of $N
$.}
\label{fig11}
\end{figure}

To address effects of the vW term in the present case, in Fig. \ref{fig11}
we plot chemical potential $\mu $ as a function of the number of atoms $N$,
using the numerical solutions of Eq. (\ref{eqqm}) with $\lambda =1/3$ and $%
\lambda =0$. It is seen that the effect of the vW gradient term is again
very weak, for any number of particles. For the same purpose, in Fig. \ref%
{fig12} we plot the radial density, $n(r,z=0)$, as found from the full
equation and produced by the TF approximation, for the same set of values of
the chemical potential as in Fig. \ref{fig10}.

\begin{figure}[tbp]
\begin{center}
{\includegraphics[width=1.02\linewidth,clip]{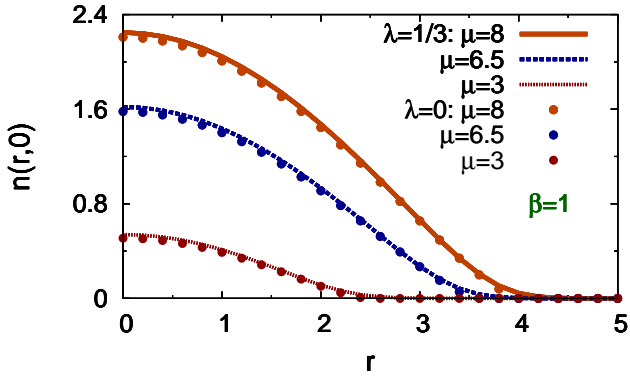}}
\end{center}
\caption{(Color online). Radial density $n(r,z=0)$ of the non-interacting
Fermi gas, with the modulation of the frequency of the transverse
confinement defined by Eq. (\protect\ref{eq_wz}), for three values of the
chemical potential: $\protect\mu =8,6.5$ and $3$. As above, symbols $\protect%
\lambda =1/3$ and $\protect\lambda =0$ designate, severally, the results
produced by the full equation (\protect\ref{eqqm}), and by the TF
approximaiton (\protect\ref{TF}). Units the same as in Fig. \protect\ref%
{fig10}.}
\label{fig12}
\end{figure}

\subsection{A self-repulsive bosonic condensate under the 2D confinement
with the strength growing in the axial direction}

A mechanism of the formation of the localized ground state, similar to that
presented in the previous section, may be applied to BEC with repulsive
interactions between atoms, subject to the action of the 2D confinement in
the plane of $\left( x,y\right) $, with the tightness growing along $z$. The
respective effective 1D equation for the mean-field wave function, $\Phi
\left( z,t\right) $, was derived in Ref. \cite{Canary}, starting from the 3D
Gross-Pitaevskii equation:%
\begin{equation}
i\hbar \frac{\partial \Phi }{\partial t}=-\frac{\hbar ^{2}}{2m}\frac{%
\partial ^{2}\Phi }{\partial z^{2}}+\hbar \omega _{\perp }(z)\sqrt{%
1+4a_{s}|\Phi |^{2}}\Phi ,  \label{DG}
\end{equation}%
where $\omega _{\perp }$ is the same trapping frequency as above [cf. Eq. (%
\ref{eqqm})], $m$ the atomic mass, $a_{s}>0$ the scattering length of the
inter-atomic interactions, and the total number of bosons is given by%
\begin{equation}
N=\int_{-\infty }^{+\infty }\left\vert \Phi (z)\right\vert ^{2}dz
\label{N1D}
\end{equation}%
(in Ref. \cite{Canary}, the norm of the 1D wave function was $1$, while $%
a_{s}$ in Eq. (\ref{DG}) was replaced by $Na_{s}$). Looking for a stationary
state as $\Phi \left( z,t\right) =a_{s}^{-1/2}\exp \left( -i\mu T\right) U(Z)
$, where $T\equiv \left( \hbar /ma_{s}^{2}\right) t$ and $Z\equiv z/a_{s}$,
we arrive at an equation for real function $U(Z)$:%
\begin{equation}
\mu U=-\frac{1}{2}\frac{d^{2}U}{dZ^{2}}+\Omega _{\perp }(Z)\sqrt{1+4U^{2}}U,
\label{Omega}
\end{equation}%
with $\Omega _{\perp }\equiv \left( ma_{s}^{2}/\hbar \right) \omega _{\perp }
$. The TF approximation for solutions of Eq. (\ref{Omega}) is obvious:%
\begin{equation}
U(Z)=\left\{
\begin{array}{c}
\left( 2\Omega _{\perp }(Z)\right) ^{-1}\sqrt{\mu ^{2}-\Omega _{\perp
}^{2}(Z)},~\mathrm{at}~\Omega _{\perp }(Z)<\mu ,~ \\
0,~\mathrm{at}~\Omega _{\perp }(Z)\geq \mu \text{,}%
\end{array}%
\right.   \label{TF1D}
\end{equation}%
cf. Eq. (\ref{TF}).

If, in particular, $\omega _{\perp }(z)$ grows faster than $\sqrt{|z|}$ at $%
|z|\rightarrow \infty $ [i.e., it obeys condition (\ref{inf})], the TF
approximation gives rise to a simple asymptotic dependence of the total
number of bosonic atoms on the chemical potential: as follows from the
substitution of expression (\ref{TF1D}) into Eq. (\ref{N1D}),%
\begin{equation}
N(\mu )=\int_{-\infty }^{+\infty }U^{2}(Z)dZ\approx \frac{\mu ^{2}}{4}%
\int_{-\infty }^{+\infty }\frac{dZ}{\Omega _{\perp }^{2}(Z)}  \label{>1/2}
\end{equation}%
for $\mu $ large enough. On the other hand, if $\omega _{\perp }$ grows
slower at $|z|\rightarrow \infty $, namely, $\Omega _{\perp }(Z)\approx
\Omega _{0}|Z|^{\alpha }$, with $\alpha \leq 1/2$, Eqs. (\ref{TF1D}) and (%
\ref{N1D}) yield, in the limit of $\mu \rightarrow \infty $:%
\begin{equation}
N(\mu )\approx \left\{
\begin{array}{c}
\left( \mu /\Omega _{0}\right) ^{2}\ln \left( \mu /\mu _{0}\right) ,~\mathrm{%
for}~\alpha =1/2, \\
\alpha \left( 1-2\alpha \right) ^{-1}\left( \mu /\Omega _{0}\right)
^{1/\alpha },~\mathrm{fort}~\alpha <1/2,%
\end{array}%
\right.   \label{<1/2}
\end{equation}%
where constant $\mu _{0}$ is determined by the structure of $\Omega _{\perp
}(Z)$ at finite values of $Z$. Note that both results (\ref{>1/2}) and (\ref%
{<1/2}) satisfy the ``anti-Vakhitov-Kolokolov" (anti-VK) criterion,
$dN/d\mu >0$, which implies the dynamical stability of the trapped
modes in the case of the self-repulsive nonlinearity \cite{HS}.

\section{Self-trapping in the self-repulsive BEC and non-interacting Fermi
gas due to the radial modulation of the 1D confinement}

The mechanism of the formation of the localized ground state in the
non-interacting Fermi gas subject to the action of the HO trapping
potential, whose strength grows from the center, can also be applied to the
case when the potential acts only in one direction, $z$, while its
tightness, $\omega _{z}^{2}$, is made a growing function of the transverse
radial coordinate, $r$. Similar to the situation considered in the previous
section, this mechanism applies not only to fermions, but also to the
self-repulsive BEC under the same type of the confinement.

In terms of such a BEC setting, a system of equations for the 2D mean-field
wave function, $\Phi \left( x,y,t\right) $, and the effective thickness of
the condensate, $\eta (x,y)$, was derived in Ref. \cite{we}:%
\begin{eqnarray}
i{\frac{\partial \Phi }{\partial t}} &=&\left[ -{\frac{1}{2}}\nabla _{\bot
}^{2}+\gamma (r)\eta ^{-1}\Phi +(1/4)\left( \eta ^{-2}+\eta ^{2}\right) %
\right] \Phi ,  \label{Phi} \\
\eta ^{4} &=&1+\gamma (r)\left\vert \Phi \right\vert ^{2}\eta ,  \label{eta}
\\
\gamma (r) &\equiv &2a_{s}\sqrt{2\pi m\omega _{z}(r)/\hbar },  \label{2D}
\end{eqnarray}%
where $a_{s}>0$ is the scattering length of the repulsive interactions
between bosonic atoms, and $\nabla _{\bot }^{2}$ is the Laplacian acting in
the plane of $\left( x,y\right) $. A straightforward analysis demonstrates
that Eqs. (\ref{Phi}) and (\ref{eta}) can be reduced to a single 2D
nonpolynomial Schr\"{o}dinger equation,%
\begin{eqnarray}
i{\frac{\partial \Phi }{\partial t}} &=&\left[ -{\frac{1}{2}}\nabla _{\bot
}^{2}+\Gamma (r)|\Phi |^{4/3}\right] \Phi ,  \label{4/3} \\
\Gamma (r) &\equiv &\frac{{5}}{4}\left( \gamma (r)\right) ^{2/3},
\label{Gamma2D}
\end{eqnarray}%
provided that the local density of the condensate is large enough:%
\begin{equation}
\gamma \left\vert \Phi \right\vert ^{2}\gg 1.  \label{cond}
\end{equation}%
Stationary solutions to Eq. (\ref{4/3}) with chemical potential $\mu $ are
looked for as%
\begin{equation}
\Phi \left( x,y,t\right) =e^{-i\mu t}U\left( r\right) ,
\end{equation}%
with real function $U(r)$ obeying the radial equation,%
\begin{equation}
\mu U=-\frac{1}{2}\left( \frac{d^{2}U}{dr^{2}}+\frac{1}{r}\frac{dU}{dr}%
\right) +\Gamma (r)U^{7/3},  \label{U}
\end{equation}%
the respective number of atoms being%
\begin{equation}
N=2\pi \int_{0}^{\infty }\left( U(r)\right) ^{2}rdr.  \label{N}
\end{equation}

Under the same conditions, a similar equation for a functional-density wave
function $U$ can be derived for the non-interacting Fermi gas\ by the
reduction of the 3D density-functional description to 2D, as shown in Ref.
\cite{David}:%
\begin{equation}
\mu U={-\nabla _{\bot }^{2}U+C_{\mathrm{2D}}}\left( {\omega _{z}(r)}\right)
^{1/3}{{{{U}^{7/3}}}},  \label{Laroze}
\end{equation}%
where $C_{\mathrm{2D}}\equiv \sqrt{{3/5}}{(6/(2s+1))^{2/3}}\pi $, and $s$ is
the semi-integer spin of the fermions. Note that, in the bosonic and
fermionic settings alike, Eqs. (\ref{4/3}), (\ref{Gamma2D}) and (\ref{Laroze}%
) demonstrate that the coefficient in front of the nonlinear term with total
power $7/3$ is proportional to $\left( {\omega _{z}(r)}\right) ^{1/3}$.

The ground state can be constructed, as above, by means of the TF
approximation, which neglects the derivatives in Eq. (\ref{U}):%
\begin{equation}
U_{\mathrm{TF}}(r)=\left[ \mu /\Gamma (r)\right] ^{3/4},  \label{mu}
\end{equation}%
hence the corresponding dependence between the chemical potential and norm (%
\ref{N}) takes a simple form,%
\begin{equation}
N_{\mathrm{TF}}=G\mu ^{3/2},~G\equiv 2\pi \int_{0}^{\infty }\frac{rdr}{\left[
\Gamma (r)\right] ^{3/2}},  \label{G}
\end{equation}%
cf. Eqs. (\ref{>1/2}) and (\ref{<1/2}). Note that the TF approximation (\ref%
{mu}) yields a fully continuous wave function (similar to that obtained by
means of the TF approximation in Ref. \cite{Barcelona}, in the model with
the spatially growing strength of the cubic self-defocusing nonlinearity in
the nonlinear Schr\"{o}dinger equation), unlike the ones found above in the
form of Eqs. (\ref{TF}) and (\ref{TF1D}), which imply divergence of the
derivatives at the boundary between the nonzero and zero parts of the
solution, $\omega _{\perp }(z)r=\sqrt{2\mu }$and $\Omega _{\perp }(Z)=\mu $,
respectively.

According to Eq. (\ref{G}), the norm of the trapped mode converges if $%
\Gamma (r)$ grows at $r\rightarrow \infty $ faster than $r^{4/3}$, i.e., as
it follows from Eqs. (\ref{Gamma2D}) and (\ref{2D}), the transverse-trapping
frequency, $\omega _{z}(r)$, must grow faster than $r^{4}$, i.e.,%
\begin{equation}
\omega _{z}(r)/r^{4}\rightarrow \infty   \label{inf2}
\end{equation}%
at $r\rightarrow \infty $. Note that the similar condition obtained in the
setting considered in the previous section was much weaker, requiring only
that $\omega _{\perp }(z)$ had to grow faster than $\sqrt{|z|}$at $%
|z|\rightarrow \infty $, see Eq. (\ref{inf}). It is also worthy to note that
the trapped TF\ modes exist with $\mu >0$, as well as solutions (\ref{TF})
and (\ref{TF1D}) considered in the previous section.

In the present notation, the condition (\ref{cond}) of the applicability of
Eq. (\ref{4/3}) for the bosonic gas takes the simple form, if TF
approximation (\ref{mu}) is used:
\begin{equation}
\mu ^{3/2}\gg 1.  \label{3/2}
\end{equation}%
This condition guarantees the applicability of Eq. (\ref{4/3}) not only to
the core of the trapped mode but also to its tail,which decays at $%
r\rightarrow \infty $, i.e., to the \emph{entire trapped mode}.

To illustrate the realization of the general setting considered in this
section, one may take the modulation function as%
\begin{equation}
\Gamma (r)=\Gamma _{0}\left( r_{0}^{2}+r^{2}\right) ^{\beta /2},
\label{beta}
\end{equation}%
with $\beta >4/3$ [this condition is necessary to meet condition (\ref{inf2}%
)]. Then, Eqs. (\ref{mu}) and (\ref{G}) yield%
\begin{eqnarray}
U_{\mathrm{TF}}(r) &=&\left( \frac{\mu }{r_{0}}\right) ^{3/4}\left(
r_{0}^{2}+r^{2}\right) ^{-3\beta /8},  \label{TF2} \\
N &=&\frac{4\pi }{3\beta -4}\left( \frac{\mu }{\Gamma _{0}r_{0}^{\beta }}%
\right) ^{3/2},  \notag
\end{eqnarray}%
cf. Eq. (\ref{>1/2}). Note that the latter $N(\mu)$ dependence also
satisfies the anti-VK criterion. Finally, it is easy to check that, in
addition to condition (\ref{3/2}) of the applicability of the underlying
equations (\ref{4/3}) and (\ref{U}), the condition for the validity of the
TF approximation (\ref{TF2}), i.e., the possibility to neglect the
derivatives in Eq. (\ref{U}) with $\Gamma (r)$ taken as per Eq. (\ref{beta}%
), amounts to $\mu \gg r_{0}^{-2}$, i.e., the trapping frequency $\omega
_{z} $ should not be too small at $r=0$.

\section{Conclusions}

In this work, we have demonstrated that the recently proposed mechanism for
the creation of bright solitons in BEC\ and nonlinear optics, by means of
the self-repulsive nonlinearity with the strength growing at $r\rightarrow
\infty $, may be also realized in Fermi gases. For the two-component
spin-balanced gas, this may be achieved by making the repulsion between the
spin components accordingly modulated in one or two directions, and the
application of the ordinary HO (harmonic-oscillator) trapping potential in
the other direction(s). The analysis is based on the Euler-Lagrange equation
produced by the minimization of the TF (Thomas-Fermi)-vW (von Weizs\"{a}%
cker) single-orbital density functional, which is quite reliable for the
description of dilute normal Fermi gases at zero temperature \cite{lipparini}%
. We have concluded\ that the vW gradient term gives nearly negligible
corrections to the TF approximations, even for relatively small numbers of
atoms. We have shown that both longitudinal and transverse widths of the
localized ground state in the Fermi gas can be efficiently controlled by
tuning the \textit{s}-wave scattering length of repulsive interactions
between spin-up and spin-down fermions, or\ by varying the number of atoms
in the ground state. Further, we have demonstrated that the localized ground
states can be created too in the non-interacting Fermi gas (in particular,
in the spin-polarized one), trapped in one of two directions by the HO
potential whose tightness grows fast enough in the remaining direction(s),
from the center to periphery. It has been also demonstrated that the latter
mechanism may create self-trapped ground states in the self-repulsive BEC
subject to the same confinement.

The analysis reported in this paper can be extended further. In particular,
it may be interesting to apply it to Bose-Fermi mixtures \cite{sala-adhi},
in the same settings which were studied here.

\section*{Acknowledgments}

LEY-S thanks FAPESP (Brazil) for partial support. LS thanks University of
Padova (progetto di ateneo 2012-2014), Cariparo Foundation (progetto di
eccellenza 2012-2014), and MIUR (progetto PRIN 2010LLKJBX) for partial
support.


\begin{thebibliography}{99}
\bibitem{chap01:njp2003b} K. E. Strecker, G.B. Partridge, A. G. Truscott,
and R. G. Hulet, 
New J.\ Phys.\ \textbf{5}, 73 (2003).

\bibitem{kono} V. A. Brazhnyi and V. V. Konotop,
Mod. Phys. Lett. B \textbf{18}, 627 
(2004).

\bibitem{fatk} F. Kh. Abdullaev, A. Gammal, A. M. Kamchatnov and L. Tomio,
Int. J. Mod. Phys. B \textbf{19}, 3415 (2005).

\bibitem{Torner} B. A. Malomed, D. Mihalache, F. Wise, and L. Torner, J.
Optics B: Quant. Semicl. Opt. \textbf{7}, R53 (2005).

\bibitem{Morsch} O. Morsch and M. Oberthaler, Rev. Mod. Phys. \textbf{78},
179 (2006).

\bibitem{extra-reviews} V. A. Yurovsky, M. Olshani, and D. S. Weiss, Adv.
At. Mol. Opt. Phys. \textbf{55}, 61 (2008); T. Lahaye, C. Menotti, L.
Santos, M. Lewenstein, and T. Pfau, Rep. Progr. Phys. \textbf{72}, 126401
(2009); Y. Kawaguchi and M. Ueda, Phys. Rep. \textbf{520}, 253 (2012).

\bibitem{boris} V. Kartashov, B. A. Malomed, and L. Torner, Rev. Mod. Phys.
\textbf{83}, 247 (2011).

\bibitem{Rosanov} N. Veretenov, Yu. Rozhdestvenskaya, N. Rosanov, V.
Smirnov, and S. Fedorov, Eur. Phys. J. D \textbf{42}, 455 (2007).

\bibitem{billam1} J. L. Helm, T. P. Billam and S. A. Gardiner, Phys. Rev. A
\textbf{85}, 053621 (2012).

\bibitem{martin} A. D. Martin and J. Ruostekoski, New J. Phys. \textbf{14},
043040 (2012).

\bibitem{Lev} B. Gertjerenken, T. P. Billam, L. Khaykovich, and C. Weiss,
Phys. Rev. A \textbf{86}, 033608 (2012).

\bibitem{Randy}   J. Cuevas, P. G. Kevrekidis, B. A. Malomed, P. Dyke, R. G.
Hulet, arXiv:1301.3959

\bibitem{billam2} T. P. Billam, A. L. Marchant, S. L. Cornish, S. A.
Gardiner, and N. G. Parker, arXiv:1209.0560.

\bibitem{sala-nl} L. Salasnich and B. A. Malomed, J. Phys. B: At. Mol. Opt.
Phys. \textbf{45}, 055302 (2012).

\bibitem{Barcelona} O. V. Borovkova, Y. V. Kartashov, B. A. Malomed, and L.
Torner, Opt. Lett. \textbf{36}, 3088 (2011).

\bibitem{Barcelona2} O. V. Borovkova, Y. V. Kartashov, L. Torner, and B. A.
Malomed, Phys. Rev. E \textbf{84}, 035602 (R) (2011); Y. V. Kartashov, V. A.
Vysloukh, L. Torner, and B. A. Malomed, Opt. Lett. \textbf{36}, 4587 (2011).

\bibitem{Zhong} W.-P. Zhong, M. Beli\'{c}, G. Assanto, B. A Malomed, and T.
Huang, Phys. Rev. A \textbf{84}, 043801 (2011).

\bibitem{Zeng} J. Zeng and B. A. Malomed, Phys. Rev. E \textbf{86}, 036607
(2012).

\bibitem{lipparini} E. Lipparini, \textit{Modern Many-particle Physics:
Atomic Gases, Quantum Dots and Quantum Fluids} (World Scientific, Singapore,
2008).

\bibitem{hkt} P. Hohenberg and W. Kohn, Phys. Rev. \textbf{136} B864 (1964).

\bibitem{tosi} N.H. March and M.P. Tosi, Ann. Phys. \textbf{81} 414 (1973);
A. Minguzzi, N.H. March, and M. Tosi, Eur. Phys. J. D \textbf{15}, 315
(2001); P. Capuzzi, Anna Minguzzi, M. P. Tosi, Phys. Rev. A \textbf{69},
053615 (2004).

\bibitem{Brandon} Brandon P. van Zyl, E. Zaremba, and P. Pisarski,
arXiv:1212.0046.

\bibitem{sala-mod} L. Salasnich, A. Cetoli, B.A. Malomed, F. Toigo, and L.
Reatto, Phys. Rev. A \textbf{76}, 013623 (2007).

\bibitem{we} L. Salasnich and B. A. Malomed, Phys. Rev. A \textbf{79},
053620 (2009).

\bibitem{Canary} A. Mu\~{n}oz Mateo and V. Delgado, Phys. Rev. \textbf{75},
063610 (2007); A \textbf{77}, 013617 (2008); A. Mu\~{n}oz Mateo, V. Delgado,
and B. A. Malomed, \textit{ibid}. \textbf{83}, 053610 (2011).

\bibitem{sala-numerics} E. Cerboneschi, R. Mannella, E. Arimondo, and L.
Salasnich, Phys. Lett. A \textbf{249}, 495 (1998); G. Mazzarella and L.
Salasnich, Phys. Lett. A \textbf{373}, 4434 (2009).

\bibitem{HS} H. Sakaguchi and B. A. Malomed, Phys. Rev. A \textbf{81},
013624 (2010).

\bibitem{David} P. D\'{\i}az, D. Laroze, I. Schmidt, and B. A. Malomed, J.
Phys. B: At. Mol. Opt. Phys. \textbf{45}, 145304 (2012).

\bibitem{sala-adhi} S. K. Adhikari and L. Salasnich, Phys. Rev. A \textbf{78}%
, 043616 (2008); S. K. Adhikari, Phys. Rev. A \textbf{72}, 053608 (2005); S.
K. Adhikari, Phys. Rev. A \textbf{70}, 043617 (2004).
\end{thebibliography}
\end{document}